\documentclass[a4paper,notitlepage,10pt]{IEEEtran}

\usepackage[utf8]{inputenc} 
\usepackage[T1]{fontenc}    
\usepackage{hyperref}       
\usepackage{url}            
\usepackage{booktabs}       
\usepackage{amsfonts}       
\usepackage{nicefrac}       
\usepackage{microtype}      
\usepackage{lipsum}		
\usepackage{graphicx}
\usepackage{mwe}
\usepackage{xcolor}
\usepackage{amsthm}
\usepackage{fontawesome}
\usepackage{comment}
\usepackage{amsmath,amssymb} 
\usepackage{color}
\usepackage{framed}
\usepackage{natbib}
\usepackage[mathscr]{euscript}
\usepackage[ruled, vlined]{algorithm2e}
\colorlet{shadecolor}{white!8}

\usepackage[a4paper,left=20mm,right=20mm,top=20mm,bottom=20mm]{geometry}

\setcitestyle{authoryear, braces=square}

\newcommand{\cycite}[2][]{\textcolor{blue}{\cite[#1]{#2}}} 

\theoremstyle{definition}
\newtheorem{definition}{Definition}[section]

\title{DeepInit Phase Retrieval}

\author{\IEEEauthorblockN{Martin Reiche} and \IEEEauthorblockN{Peter Jung}}

\date{}

\begin{document}
\maketitle

\begin{abstract}

  This paper shows how data-driven deep generative models can be
  utilized to solve challenging \emph{phase retrieval} problems, in
  which one wants to reconstruct a signal from only few intensity
  measurements. Classical iterative algorithms are known to work well
  if initialized close to the optimum but otherwise suffer from
  non-convexity and often get stuck in local minima. We therefore
  propose \emph{DeepInit Phase Retrieval}, which uses regularized
  gradient descent under a deep generative data prior to compute a
  \emph{trained} initialization for a fast classical algorithm
  (e.g. the randomized Kaczmarz method).  We empirically show that our
  hybrid approach is able to deliver \emph{very high} reconstruction
  results at low sampling rates even when there is significant
  generator model error.
  Conceptually, learned initializations may therefore help to overcome
  the non-convexity of the problem by starting classical descent steps
  closer to the global optimum. Also, our idea demonstrates superior
  runtime performance over conventional gradient-based reconstruction
  methods. We evaluate our method for generic measurements and show
  empirically that it is also applicable to diffraction-type
  measurement models which are found in terahertz single-pixel phase
  retrieval.

\end{abstract}

\section{Introduction}
\label{chapter:introduction}

\label{section:introduction:nonlinear_inverse_problems_real_world_applications}

An important nonlinear inverse problem is the \emph{phase retrieval}
problem, which has been extensively studied over the last decades
because it arises in a lot of applications,
e.g. crystallography~\cycite{harrison_phase_1993}\cycite{millane_phase_1990},
astronomy~\cycite{fienup_phase_1987} and (optical) imaging
\cycite{shechtman_phase_2014}. It was originally formulated as the
problem of reconstructing a signal from the magnitude of its Fourier
transform which, for example, arises when illuminating a scene with a
coherent electromagnetic field and measuring its magnitude
in the far field. A classical alternating recovery
approach here is the well-known \emph{Fienup algorithm}~\cycite{fienup_phase_1982}
which iteratively imposes real-plane and Fourier-plane constraints.

A related variant of the aforementioned problem is the reconstruction
of the scene from its diffraction pattern. Diffraction occurs when a
wave hits an obstacle (whose size is in the order of the wavelength)
or passes an aperture, and therefore influences its complex-valued
wave pattern.  Since most detectors can only measure intensity, phase
retrieval techniques are necessary to recover the original signal.
For example, in single-detector imaging a target scene is repeatedly
illuminated with radiation that has passed a spatial light modulator
configured with a random on/off pixel pattern. The modulated radiation
pattern then hits the scene and its transmission is collected through
a collecting optics (e.g. a lens) at a single detector cell.
With an accurate forward diffraction model
it is possible to computationally recover the scene from the collected
intensity measurements (computational imaging).

An important aspect to improve recovery and reduce acquisition time is
to incorporate prior knowledge about targets in the scene. For
example, an intuitive assumption is that the signals may be
(approximately) represented as sparse vectors in some known transform
domain. However, in many applications sparsity is often a too simple
description of realistic signal structures. Therefore, a more recent approach is
to learn the signal structure directly from training data. Neural
networks can be trained to denoise desired signals while
\emph{generative neural networks} can learn a particular source
characteristics. Not surprisingly, a lot of prior research has been
conducted at the intersection of deep learning and nonlinear inverse
problems.

In a more generic formulation of phase retrieval, one wants to find an
$n$-dimensional signal vector $\mathbf{x}$ such that the vector
$|\mathbf{A}\mathbf{x}|^2$ of intensities is consistent with an
observation vector $\mathbf{y}\in\mathbb{R}_+^m$ where
$\mathbf{A} \in \mathbb{C}^{m \times n}$ is a given complex-valued
measurement (or sensing) matrix (modeling the wave propagation in the diffraction imaging example mentioned above). In practice one always has to
consider measurement noise due to non-optimal detectors and then a
particularly simple approach is to formulate recovery for example as:
\begin{equation}
  \label{eq:inverse_problems:phase_retrieval}
  \min_{\mathbf{x}} \Vert \mathbf{y} - |\mathbf{Ax}|^2 \Vert^2_2
\end{equation}
Given a more concrete noise model, other loss functions are also
feasible for applications.  A fundamental question however is when a
signal is uniquely (up to trivial ambiguities) determined by noiseless
intensity measurements.
For example, Eldar and Mendelson have shown that for
$m=\mathcal{O}(n)$ subgaussian measurements the problem has with
overwhelming probability a unique (up to its sign) solution
\cycite{eldar_phase_2014}. The precise scaling has a longer history,
see for example \cycite{bandeira_saving_2014}.  Further uniqueness
results exist, e.g., in the case of random binary matrices, see
\cycite{krahmer_phase_2018,krahmer_complex_2019}.

Besides identifiability it is important for applications to solve the
problem also with robust and computationally tractable algorithms
obeying rigorous guarantees.  Fienup's algorithm is known to be very
efficient but is not guaranteed to recover the correct solution
\cycite{osherovich_numerical_2012}. On the other hand, semidefinite
relaxations like \emph{PhaseLift}~\cycite{candes_phaselift:_2011}
yield a convex problem for which rigorous guarantees
exists. But such lifting approaches are extremely computationally
demanding and are therefore more of theoretical than of practical use.

To overcome computational burden, gradient descent based approaches
for the nonconvex loss function have been investigated intensively.
For example, Candes et al. proposed a \emph{Wirtinger Flow}
approach~\cycite{candes_phase_2015} that aims to minimize the
intensity loss $\|\mathbf{y} - |\mathbf{A}\mathbf{x}|^2\|^2_2$.  This
method was extended to use truncated
gradients~\cycite{chen_solving_2015} which converge faster to the
optimal value.  Wang et al. proposed \emph{Truncated Amplitude
  Flow}~\cycite{wang_solving_2016} which minimizes the amplitude loss
$\|\sqrt{\mathbf{y}} - |\mathbf{A}\mathbf{x}|\|^2_2$. Especially for
real-valued signals, Tan and Vershynin proposed a very fast
\emph{randomized Kaczmarz} approach by iteratively choosing one of the
measurements
$\sqrt{y_i} = |\langle \mathbf{a}_i, \mathbf{x} \rangle| $ at random
and projecting onto the closer of the two hyperplanes corresponding to
$\pm\mathbf{x}$ \cycite{tan_phase_2017}.

However, since all these iterative approaches operate on non-convex
loss functions, careful initialization is necessary in
practice. In real world imaging applications it is important to have an
initialization which is also close to the optimum so that descent
algorithms run into the correct local minima. This is 
exactly a point where learning may help in classical algorithms.
For example, autoencoders can be used as trainable denoisers
to improve reconstruction when used as regularizers or proximal
mappings in iterative algorithms. For example,
\emph{prDeep}~\cycite{metzler_prdeep:_2018} is based on the
regularization by denoising (RED)
approach~\cycite{romano_little_2017} to minimize the amplitude-based
objective function by adding a generative neural network-based
denoiser regularization term (they used the well-known
\emph{DnCNN}~\cycite{zhang_beyond_2017} network).

\section{DeepInit Phase Retrieval}
\label{chapter:nonlinear_inverse_problems_deep_generative_models}
    
As indicated above, prior information about permissible signals $\mathbf{x}$ may
drastically improve the reconstruction for  phase retrieval
algorithms. In particular, generative models based on deep
(feed-forward) neural networks are interesting as they are able to learn
to generate samples even from very complicated signal distributions
(such as e.g. natural images~\cycite{gulrajani_pixelvae:_2016} or
faces~\cycite{karras_style-based_2018}).

Using a deep generative model as a prior follows the idea that we have
given a generator $G: \mathbb{R}^p \rightarrow \mathbb{R}^n$ for
$p\ll n$ that has been trained to sufficiently well generate
permissible signals.  Instead of reconstructing a signal
$\mathbf{x}^*$ directly, one considers then (in the case of intensity
loss as data fidelity):
\begin{equation}
  \label{eq:solving:deep_generative_models_priors_objective}
  \min_{\mathbf{z}}
  \|\mathbf{y}- |\mathbf{A}G(\mathbf{z})|^2\|_2^2+\lambda R(\mathbf{z})
\end{equation}
where $R:\mathbb{R}^p\rightarrow\mathbb{R}_+$ is a given regularizer
function. Obviously, due to the quadratic measurements and the nature
of $G$ the objective above is non-convex and can usually minimized
only locally using descent methods and a good initialization. A point
$\mathbf{z}^*$ in the latent space obtained in this way yields then a
reconstruction $G(\mathbf{z}^*) = \mathbf{x}^*$ (same for other loss
functions, like the amplitude loss).  In the case of linear
measurements and for $\lambda=0$, this approach has been investigated
in \cycite{bora_compressed_2017} as a data-driven extension of compressed sensing.

In \cycite{hand_phase_2018} it has been proven that for differentiable
generator networks consisting of layers with ReLU activation functions
and Gaussian $\mathbf{A}$, the unregularized ($\lambda=0$) objective
function in \eqref{eq:solving:deep_generative_models_priors_objective}
does (with overwhelming probability) not have spurious local minima away
from neighborhoods of the true solution (or negative multiples
thereof).  An analogous proof has been provided for the linear case
already in \cycite{hand_global_2017} including a Tikhonov
regularization term $R(\mathbf{z})=\|G(\mathbf{z})\|^2_2$.
Asim et al. experimentally validated an approach to solve a
regularized linear case with
$R(\mathbf{z})=\Vert \mathbf{z}\Vert_2$ using L-BFGS with
great success on important computer vision datasets where $G$ was
trained to operate on normal distribution in the latent space
\cycite{asim_invertible_2019}.
Further works by \cycite{shamshad_robust_2018} and
\cycite{hand_phase_2018} propose to minimize the (unregularized)
amplitude loss
$\Vert \sqrt{\mathbf{y}} - | \mathbf{A}G(\mathbf{z}) | \Vert^2$ with a
gradient descent scheme.

In this work we will consider TV-regularized phase retrieval with
$R(\mathbf{z})=\|G(\mathbf{z}) \|_{\mathrm{TV}}$ in
\eqref{eq:solving:deep_generative_models_priors_objective}, i.e., for
a given generator $G$ the problem is:
\begin{equation}
  \label{eq:solving:drgd}
  \min_{\mathbf{z}} \Vert \mathbf{y} - |\mathbf{A}G(\mathbf{z})|^2 \Vert^2_2 + \lambda \Vert G(\mathbf{z}) \Vert_{\mathrm{TV}}
\end{equation}
where $\|\cdot\|_{\mathrm{TV}}$ is the \emph{discrete anisotropic
  total variation norm}.  In the linear case such a regularized
approach has been investigated already in \cycite{Ulyanov2017} and
with additionally learned regularizers in \cycite{VanVeen2018}.
Formulation \eqref{eq:solving:drgd} is especially beneficial when the
generator has not been properly trained and therefore yields notable
residual model error $\min_\mathbf{z}\|\mathbf{x} - G(\mathbf{z})\|$.
Problem~\eqref{eq:solving:drgd} can be solved using (sub-)gradient
descent. However, since the optimization \eqref{eq:solving:drgd} in
terms of $\mathbf{x}=G(\mathbf{z})$ is limited to the range of the
generator, the achieved performance will depend critically on the
quality of $G$ and how representative the training data is for the
target to recover.

\subsection*{Deep Generative Initialization}
\label{section:nonlinear_inverse_problems_deep_generative_models:deep_generative_initialization}
    
Given the shortcomings of the deep generative prior-based
reconstruction methods described above with respect to the generator's
model error, we propose to use a hybrid approach that takes the
reconstruction result of a generative prior-based method and uses that
as initialization for a classical algorithm.  Conceptually, this works
as follows: First we reconstruct an approximate
$\tilde{\mathbf{x}}= G(\tilde{\mathbf{z}})$ using a randomly
initialized sub-gradient descent for problem
\eqref{eq:solving:drgd}.  We call
this step ``Deep Regularized Gradient Descent'' (DRGD).  As we have
discussed, the similarity of $\tilde{\mathbf{x}}$ to the true
$\mathbf{x}$ is depending on the generator $G$'s ability to correctly
model the signal domain. Then we use
$\tilde{\mathbf{x}}= G(\tilde{\mathbf{z}})$ as initialization
$\mathbf{x}^{(0)}$ for a traditional reconstruction method in order
to solve \eqref{eq:inverse_problems:phase_retrieval}.  Doing this, we
overcome the model error of the generator $G$ but are still able to
use the prior knowledge encoded into it without compromising
reconstruction quality.
    
\begin{figure}
      \centering
      \includegraphics[width=1\linewidth]{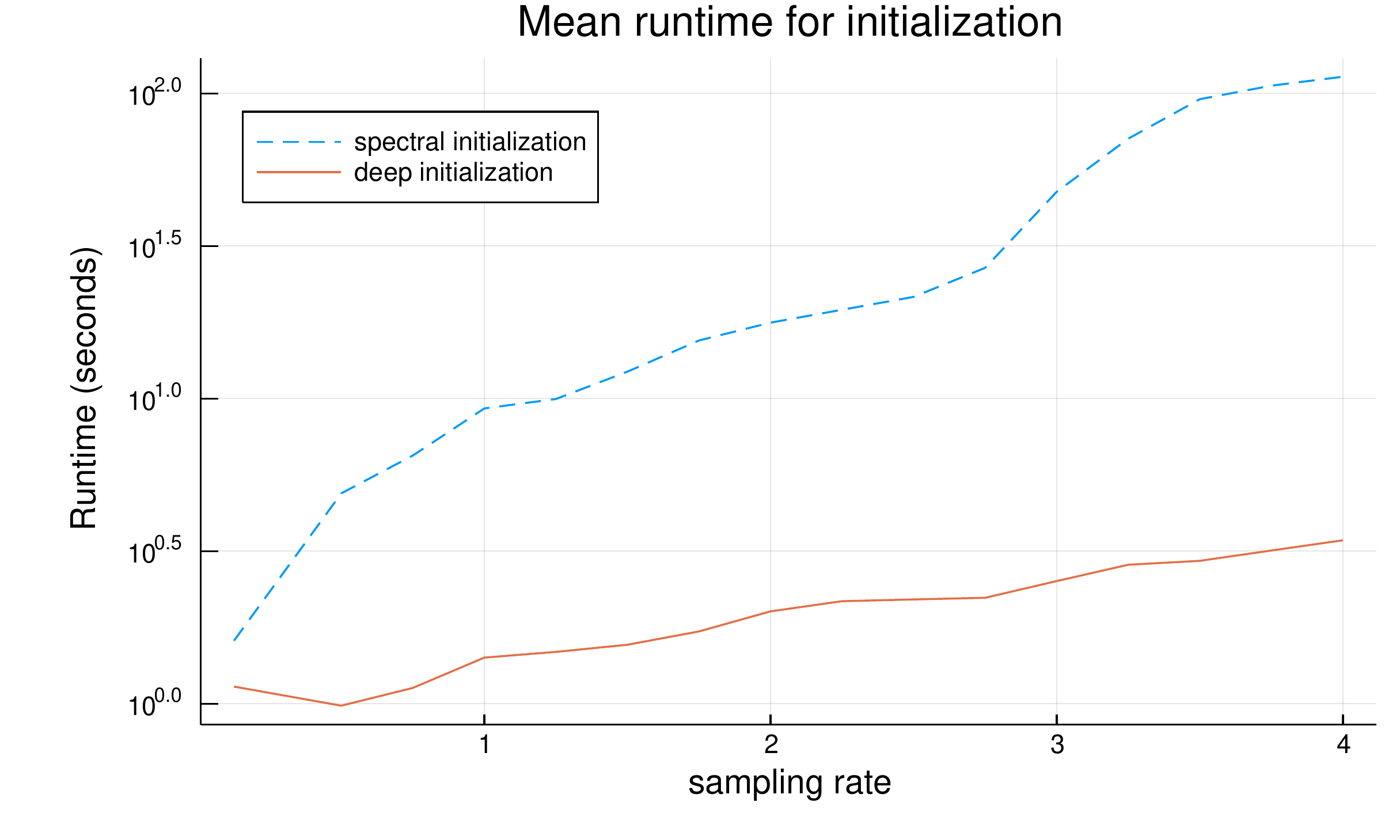}
      \caption{Comparison of the runtime of  spectral initialization against deep generative initialization. Note the logarithmic scale on the vertical axis. All results are averaged over five different images.}
      \label{fig:results:spectral_vs_deep_initialization}
    \end{figure}

    Our hypothesis is that based on this data-driven initialization
    $\mathbf{x}^{(0)}$ of the reconstruction method, we get better
    reconstruction results as we start closer to the true value of
    $\mathbf{x}$. We furthermore need potentially less iterations to
    get to comparable reconstruction errors.  In comparison to the
    spectral initialization method employed by most reconstruction
    methods, our initialization method is also significantly
    faster. Figure~\ref{fig:results:spectral_vs_deep_initialization}
    shows comparisons of both initialization methods with respect to
    runtime.
    
    Below, we investigate our hybrid approach exemplary for the
    combination of the Deep Regularized Gradient Descent (DRGD) for
    \eqref{eq:solving:drgd} and the Randomized Kaczmarz (RK) algorithm
    for \eqref{eq:inverse_problems:phase_retrieval},
    which we will name {\em DeepInit Phase Retrieval} (DeepInit).  The
    pseudocode is shown in Algorithm~\ref{algorithm:drgd_rk}.  We will
    evaluate its performance in
    Section~\ref{section:solving:numerical_experiments}.
    
    \begin{algorithm}
    
     \KwData{measurements $\mathbf{y} = \left[ y_1,...,y_m \right]^\top \in \mathbb{R}^{m}$
     	\newline sensing matrix $\mathbf{A} = \left[ \mathbf{a}_1,...,\mathbf{a}_m \right]^\top \in \mathbb{C}^{m \times n}$
     	\newline differentiable generator network \newline \noindent\hspace*{2mm} $G(\mathbf{z}): \mathbb{R}^p \rightarrow \mathbb{R}^n$
     	\newline step size $\eta$
     	\newline regularization parameter $\lambda$
     	\newline number of iterations of the initializer $i_{\mathrm{max}}$
        \newline number of iterations of the randomized Kaczmarz method $k_{\mathrm{max}}$
        }
     \KwResult{
        reconstruction $\mathbf{x}^{(k_{\mathrm{max}})}$
     }
     Randomly initialize $\mathbf{z}^{(0)} \in \mathbb{R}^p$
     
     \SetAlgoLined
     \For{$i = 0$ \KwTo $i_{\mathrm{max}}-1$}{
        $\mathbf{z}^{(i+1)} \leftarrow \mathbf{z}^{(i)} - \eta \nabla_{\mathbf{z}^{(i)}} \left(\Vert \mathbf{y} - |\mathbf{A}G({\mathbf{z}^{(i)}})|^2 \Vert^2_2 + \lambda \Vert G({\mathbf{z}^{(i)}}) \Vert_{\mathrm{TV}} \right) $
     }
     
     $\mathbf{x}^{(0)} \leftarrow G(\mathbf{z}^{(i_{\mathrm{max}})})$
     
     \SetAlgoLined
     \For{$k = 0$ \KwTo $k_{\mathrm{max}}-1$}{
        $\mathbf{x}^{(k+1)} \leftarrow \mathbf{x}^{(k)} + \left( \frac{ \mathrm{sign}(\langle \mathbf{a}_{r(k+1)}, \mathbf{x}^{(k)} \rangle) \sqrt{y_{r(k+1)}} - \langle \mathbf{a}_{r(k+1)}, \mathbf{x}^{(k)} \rangle }{\Vert \mathbf{a}_{r(k+1)} \Vert^2_2} \right)\mathbf{a}_{r(k+1)}$
     }
     
     \caption{DeepInit Phase Retrieval algorithm} 
     \label{algorithm:drgd_rk}
    \end{algorithm}


\section{Numerical Experiments}
\label{section:solving:numerical_experiments}
    
In this section we will present a numerical evaluation of the proposed
methods for complex Gaussian, i.e., unstructured measurements.  We compare our
approach against traditional phase retrieval methods (Wirtinger Flow
(WF), Truncated Wirtinger Flow (TWF), and Randomized Kaczmarz (RK)) on
one standard test dataset and one synthetically generated dataset
using structural similarity index (SSIM) and peak signal-to-noise
ratio (PSNR), two important quality metrics commonly used for image
quality evaluation tasks.

\subsection{MNIST dataset}
\label{section:solving:mnist_dataset}
\begin{figure}
            \centering
            \includegraphics[width=1.0\linewidth]{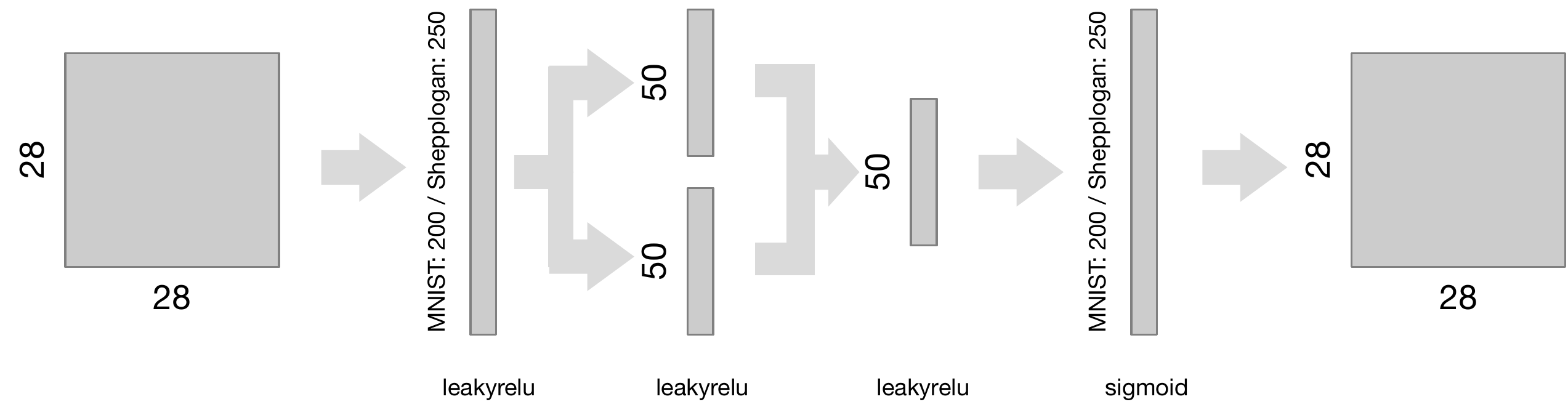}
            \caption{Architecture of the variational autoencoder trained on MNIST and the Shepp-Logan dataset.}
            \label{fig:numerical_experiments:mnist_vae}
        \end{figure}

We evaluate the performance of the reconstruction methods on the
well-known MNIST dataset~\cycite{lecun_mnist_1998} consisting of
$60,000$ handwritten digits represented as $28 \times 28$ pixel
grayscale
images.
For DRGD and DeepInit, we train a variational autoencoder on this
dataset.
        The architecture of the model can be seen in Figure \ref{fig:numerical_experiments:mnist_vae}. We use an $\ell^2$-regularized ELBO as the loss function and train for 50 epochs in batches of 128 images.
        
\subsection{Shepp-Logan dataset}
\label{section:solving:shepplogan_dataset}
\begin{figure}
            \centering
            \includegraphics[width=1\linewidth]{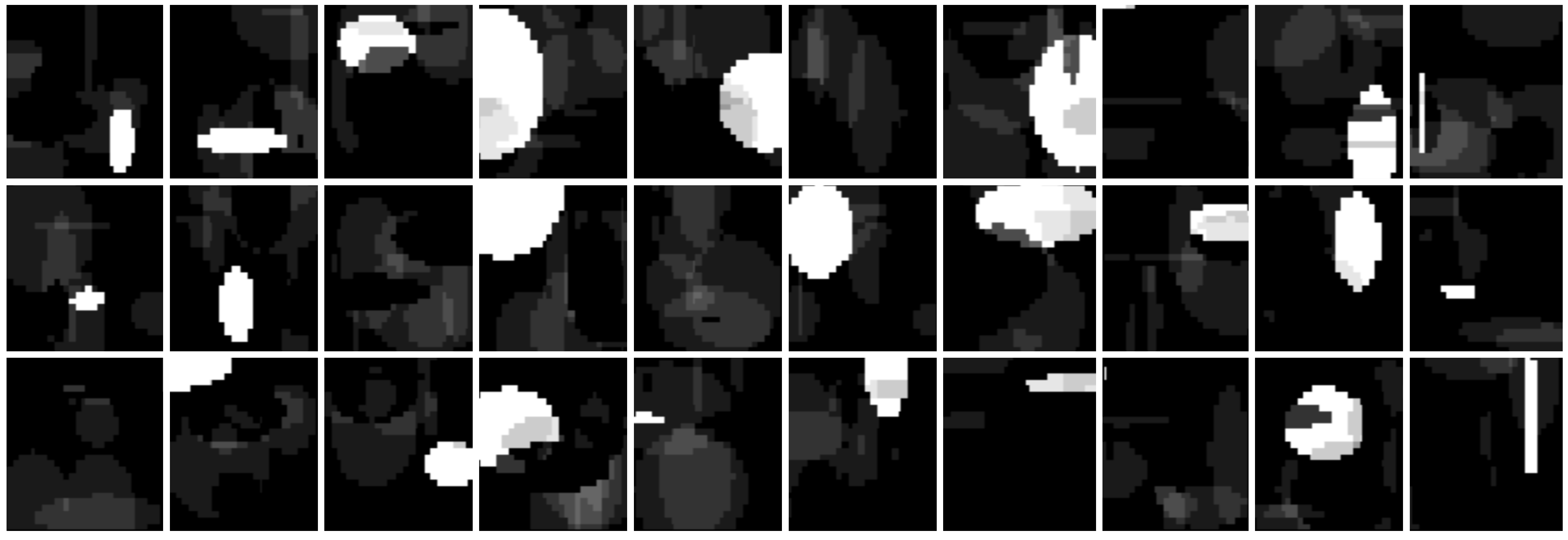}
            \caption{30 samples taken from the synthetically generated Shepp-Logan dataset.}
            \label{fig:numerical_experiments:shepplogan_kaleidoscope}
    \end{figure}

    This synthetically generated dataset is inspired by the well-known Shepp-Logan phantom by randomizing the location and size parameters of its constituent shapes. As in the MNIST dataset, each image of the Shepp-Logan dataset is $28 \times 28$ pixels in size and grayscale. The overall dataset (see Figure~\ref{fig:numerical_experiments:shepplogan_kaleidoscope}) contains $250,000$ randomly generated images.
    
    Analog to the MNIST dataset, we train a variational autoencoder using the Shepp-Logan dataset with the same loss function, batch size and epoch count that we used for the MNIST dataset. We adapt the network design of the variational autoencoder slightly as shown in Figure~\ref{fig:numerical_experiments:mnist_vae}.


\subsection{Numerical Experiments}

All evaluations are performed under a noise-free measurement model
$\mathbf{y} = |\mathbf{Ax}|^2$ using complex random Gaussian
measurement matrices $\mathbf{A}$.
We evaluate the methods based on
their reconstruction quality and runtime and set their parameters as follows:
Wirtinger Flow (WF) uses $k_\mathrm{max} = 50$ iterations;
Truncated Wirtinger Flow (TWF) uses $k_\mathrm{max} = 200$
iterations; Randomized Kaczmarz (RK) uses
$k_\mathrm{max} = 100000$ iterations;\footnote{Note that
  Randomized Kaczmarz, unlike Wirtinger Flow or
  Truncated Wirtinger Flow, is not a gradient-based method
  and that iteration counts therefore are not comparable. In
  general, Randomized Kaczmarz iterations are much faster
  to execute than gradient steps.} Deep Regularized Gradient
    Descent (DRGD) uses $k_\mathrm{max} = 200$ iterations with step size
    $\eta = 0.1$ and regularization factor $\lambda = 0.1$;
    DeepInit Phase Retrieval uses $i_\mathrm{max} = 200$ iterations for the initializer with step size $\eta = 0.1$ and regularization factor $\lambda = 0.1$, and $k_\mathrm{max} = 100000$ as the iteration count for the Randomized Kaczmarz part.
    
\begin{figure}
        \centering
        \includegraphics[width=1\linewidth]{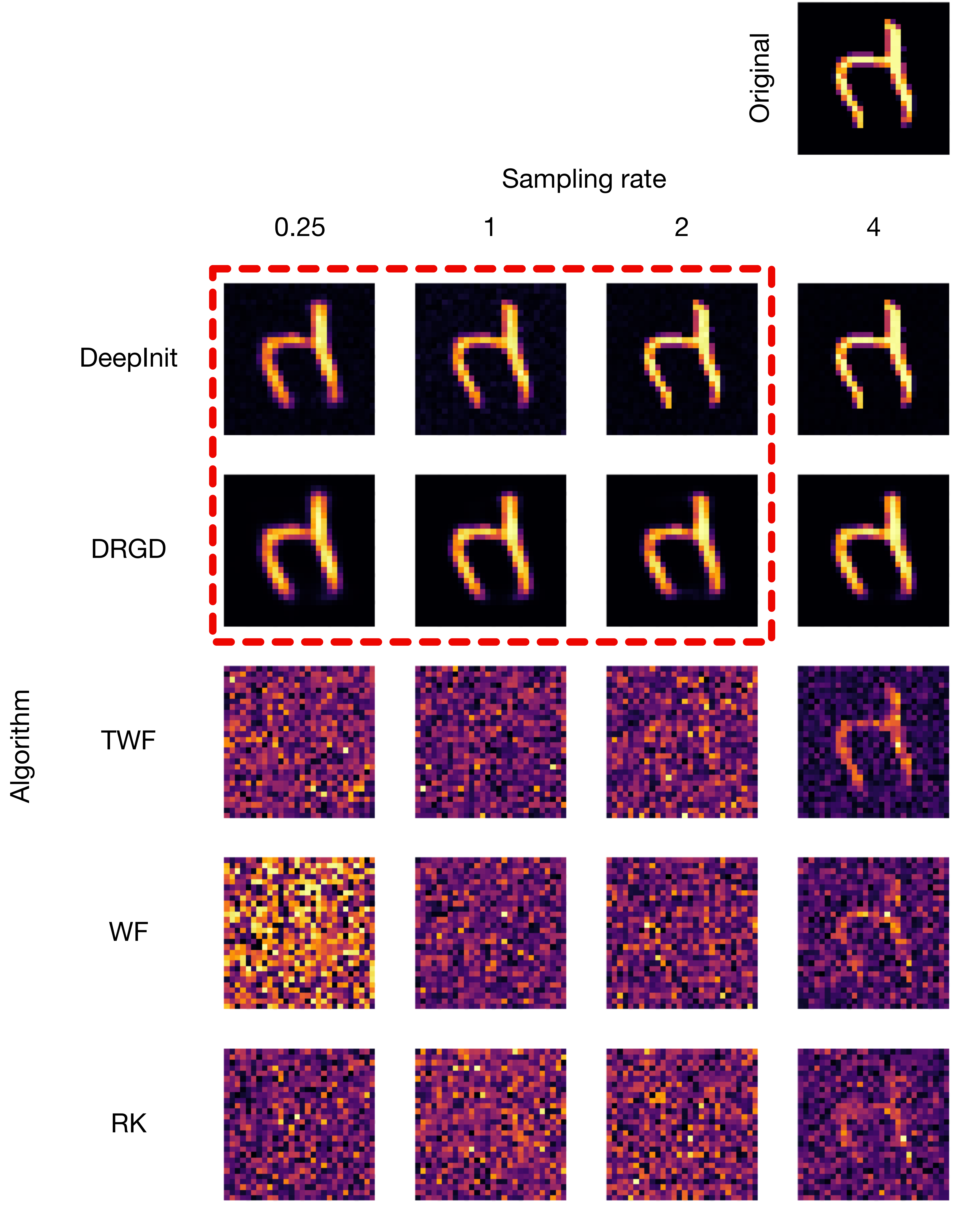}
        \caption{Results of the reconstruction process for a selected MNIST test image for all algorithms and selected sampling rates. Deep Regularized Gradient Descent and DeepInit Phase Retrieval make use of a trained variational autoencoder. Important results are highlighted with a dashed red box.}
        \label{fig:results:main:visual:mnist}
    \end{figure}

\begin{figure}
        \centering
        \includegraphics[width=1\linewidth]{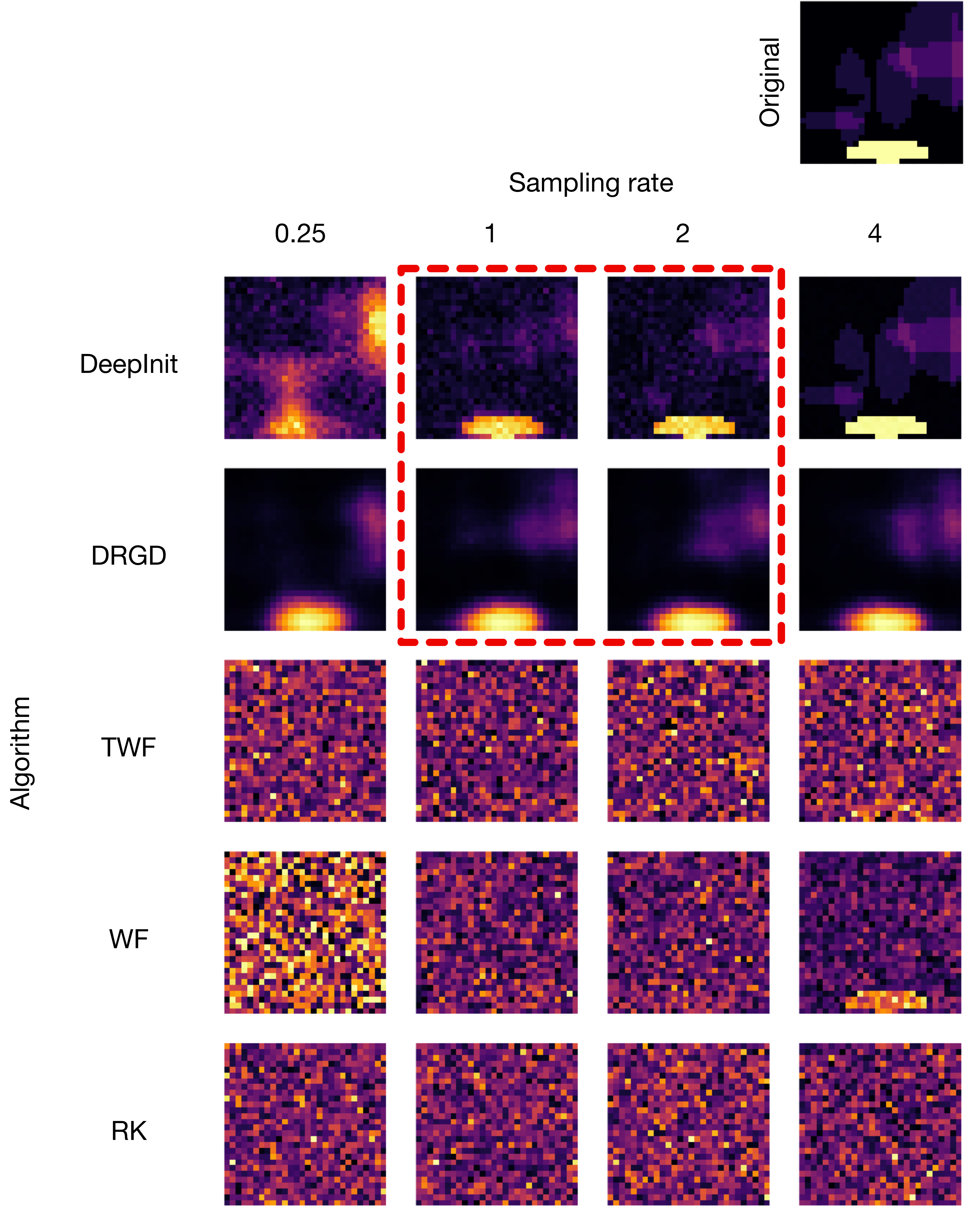}
        \caption{Results of the reconstruction process for a selected Shepp-Logan test image for all algorithms and selected sampling rates. Deep Regularized Gradient Descent and DeepInit Phase Retrieval make use of a trained variational autoencoder. One can see that Deep Regularized Gradient Descent is unable to perform a good reconstruction due to an inability of the variational autoencoder to model the original signal distribution well enough. Important results are highlighted with a dashed red box.}
        \label{fig:results:main:visual:shepplogan}
    \end{figure}

\begin{figure}
        \centering
        \includegraphics[width=1\linewidth]{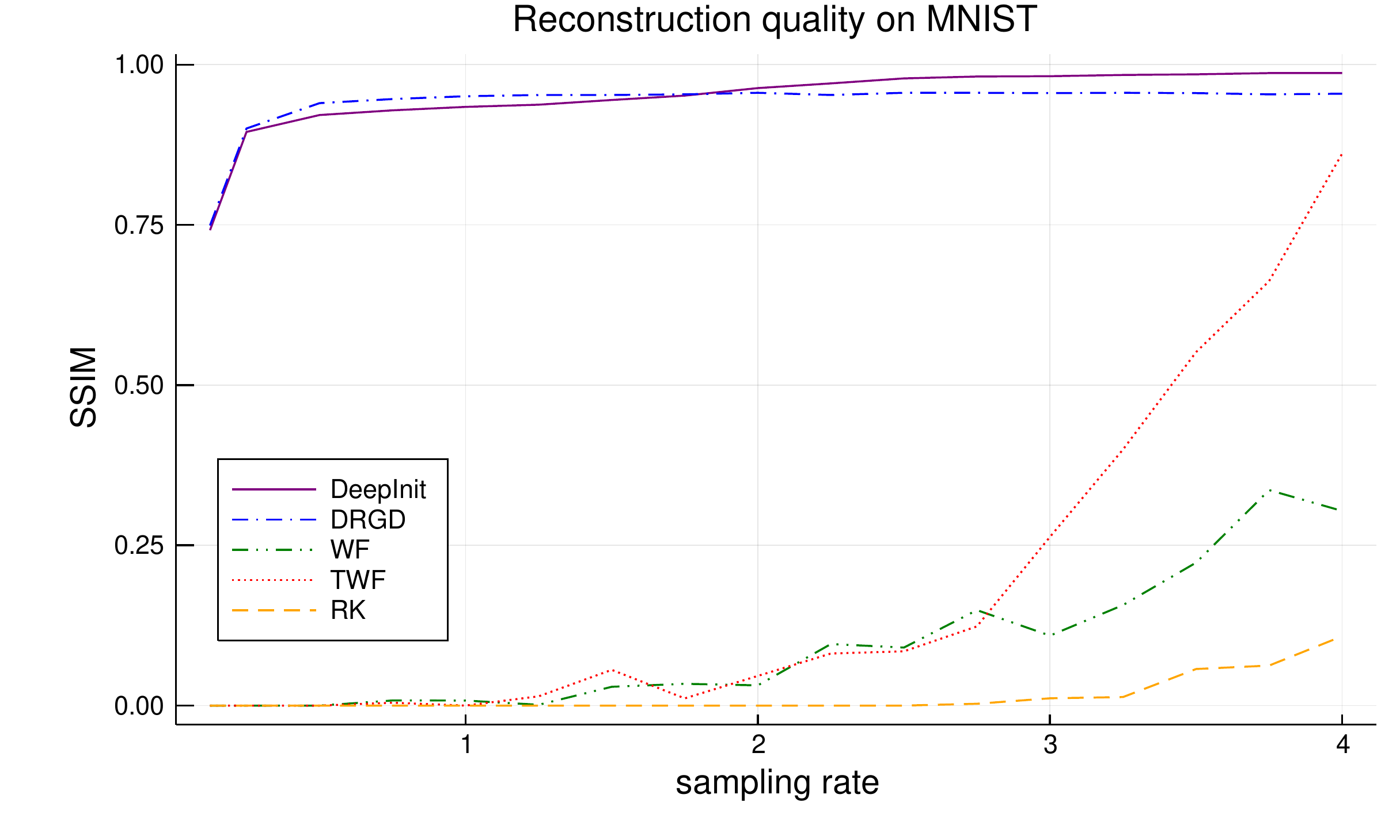} \\
        \includegraphics[width=1\linewidth]{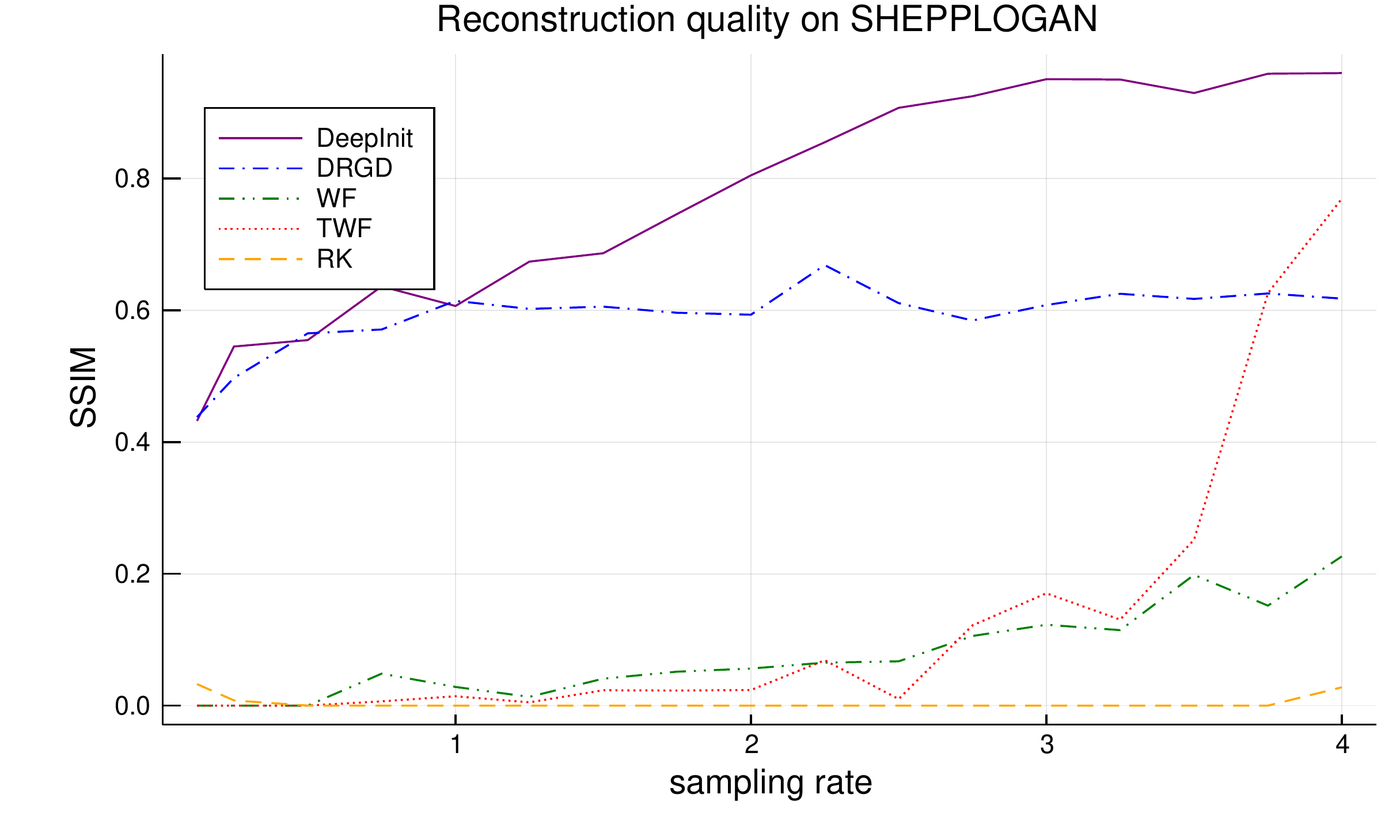}
        \caption{Evaluation results for the MNIST and Shepp-Logan datasets with respect to the structural similarity index for image quality. All results are averaged over five different images.
        Note that the reconstruction quality of DeepInit Phase Retrieval is upper-bounded by a fixed number of Kaczmarz iterations.}
        \label{fig:results:main:ssims}
    \end{figure}

\begin{figure}
        \centering
        \includegraphics[width=1\linewidth]{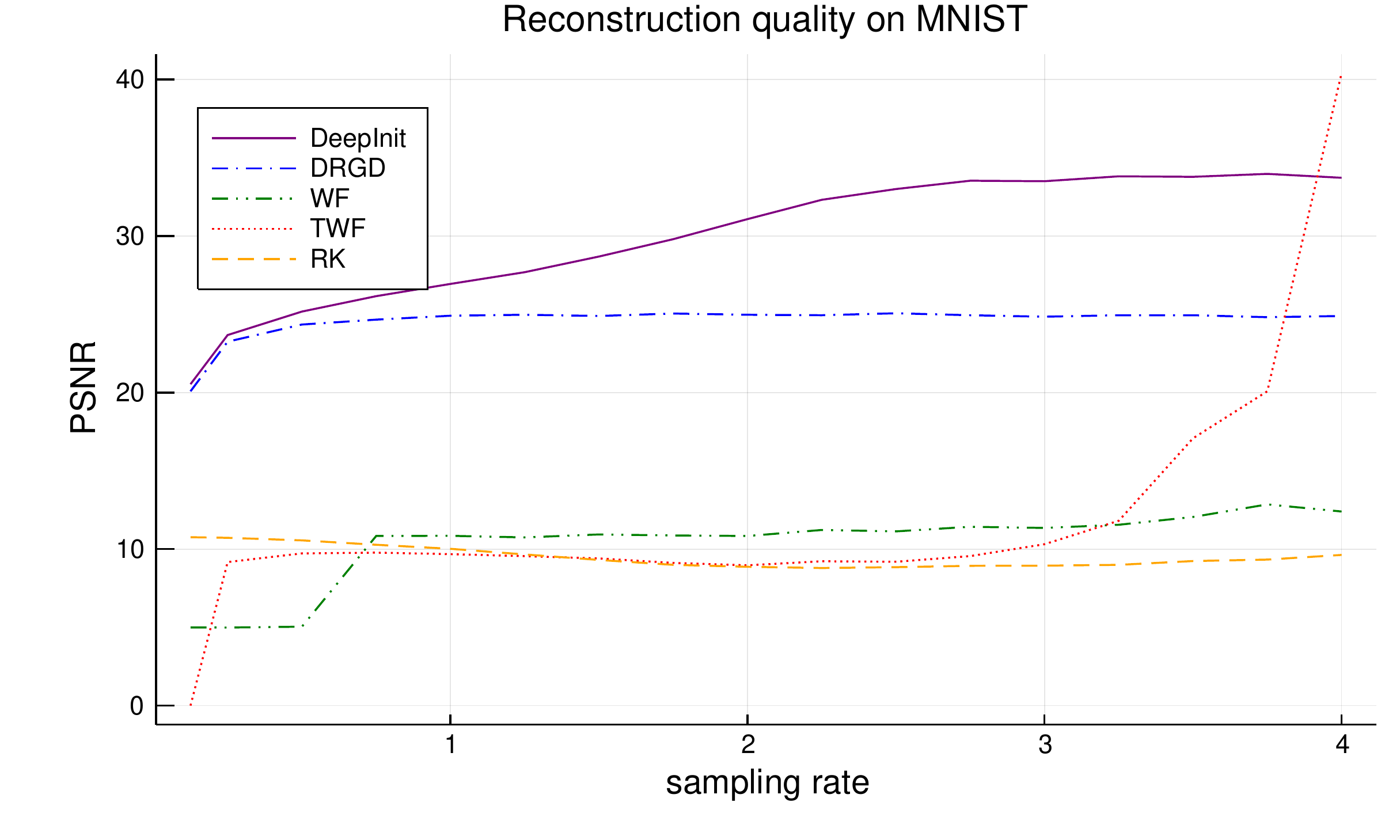} \\
        \includegraphics[width=1\linewidth]{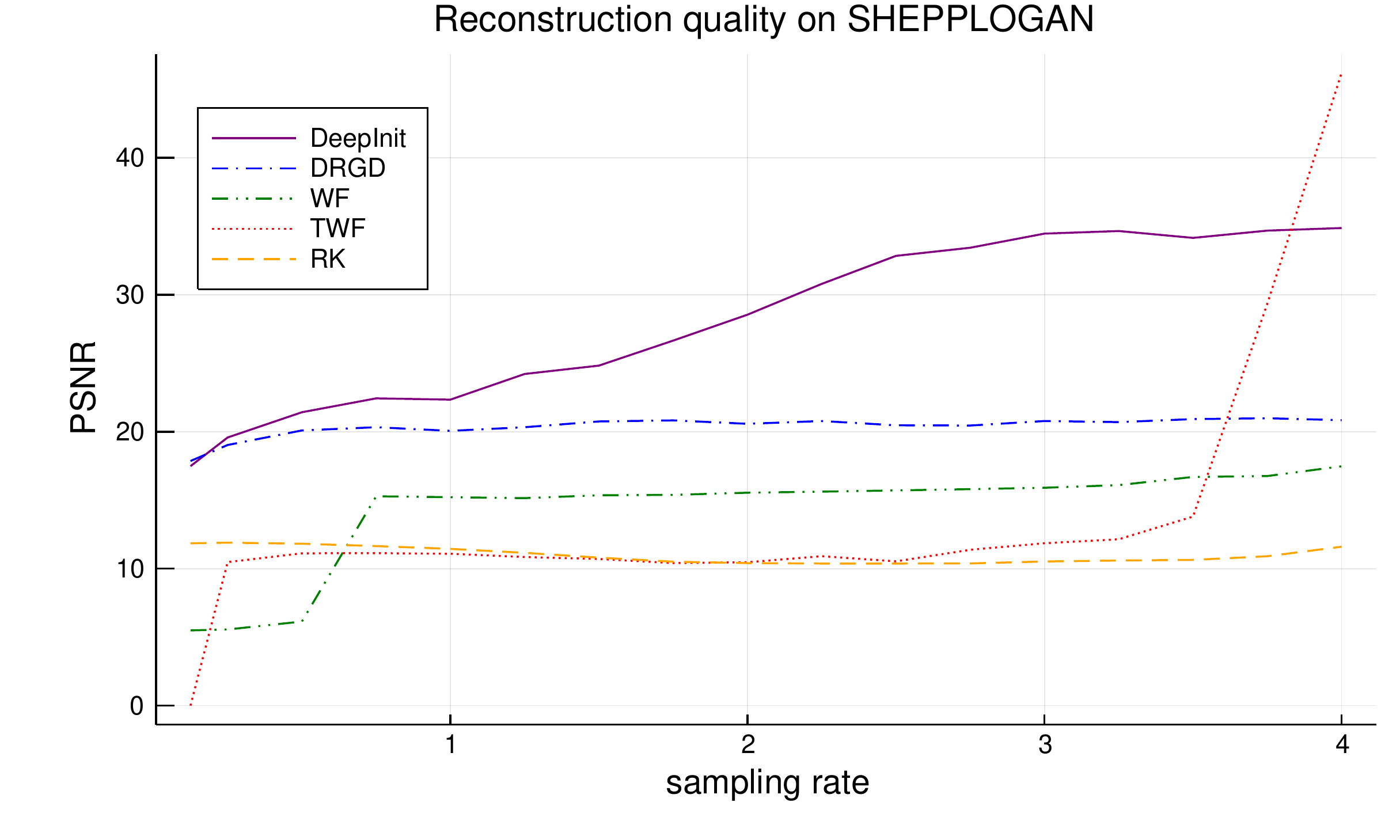}
        \caption{Evaluation results for the MNIST and Shepp-Logan datasets with respect to the peak signal-to-noise ratio. All results are averaged over five different images.
        Note that the reconstruction quality of DeepInit Phase Retrieval is upper-bounded by a fixed number of Kaczmarz iterations, which is visible by the quality plateau that the algorithm reaches.}
        \label{fig:results:main:psnrs}
    \end{figure}

    Figures~\ref{fig:results:main:visual:mnist} and
    \ref{fig:results:main:visual:shepplogan} visually show the
    reconstruction results for selected images from the MNIST and
    Shepp-Logan datasets and highlight the most important
    results. Figures~\ref{fig:results:main:ssims} to
    \ref{fig:results:main:psnrs} show the evaluation results with
    respect to reconstruction quality under SSIM and PSNR.

    \begin{figure}
        \centering
        \includegraphics[width=1\linewidth]{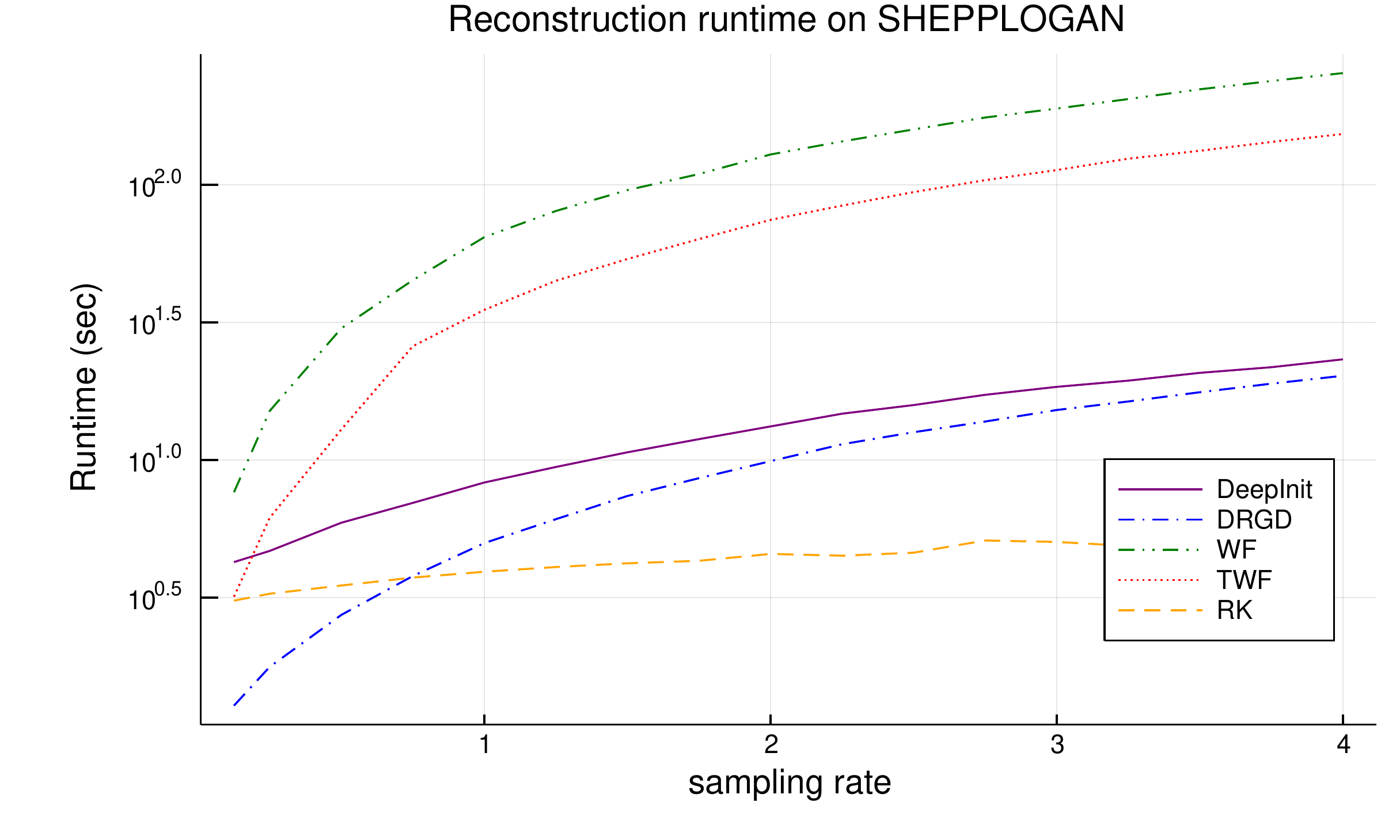}
        \caption{Evaluation results for the MNIST and Shepp-Logan datasets with respect to the reconstruction time. Note the log scale of the vertical axis. All results are averaged over five different images.}
        \label{fig:results:main:runtime}
    \end{figure}

\subsection{Main findings and observations}
    
    \subsubsection{Reconstruction quality}
    
    On can see that for the evaluated sampling rate range (0.125 to 4.0), DeepInit Phase Retrieval and Deep Regularized Gradient Descent show superior reconstruction quality over Truncated Wirtinger Flow, Wirtinger Flow and Randomized Kaczmarz for both evaluation data sets.
    
    At a sampling rate of 4.0, the reconstruction quality of DeepInit Phase Retrieval is marginally lower compared to Truncated Wirtinger Flow. This effect can be attributed to the fixed number of randomized Kaczmarz iterations in the experiment and is especially visible in the results with respect to PSNR.
    
    Because it only allows solutions that lie in the range of the generator, Deep Regularized Gradient Descent fails to deliver competitive reconstruction results 
    due to an inability of the variational autoencoder to model the original signal distribution well enough. This effect is not visible in reconstructions using the DeepInit Phase Retrieval algorithm, because in this method the final solution is not bound to lie in the range of the generator network.
    
    \subsubsection{Runtime}
    Our evaluation shows that DeepInit Phase Retrieval and Deep Regularized Gradient Descent both have drastically superior runtime performance when compared to Truncated Wirtinger Flow or Wirtinger Flow (see Figure~\ref{fig:results:main:runtime}).

\section{DeepInit Phase Retrieval for Diffraction Single-Pixel Imaging}
\label{chapter:practical_application}
    
So far we have discussed deeply initialized phase retrieval algorithms
from a generic point of view for complex Gaussian intensity
measurements. In this section we shall consider now a practical
application of the proposed algorithms for {\em single-pixel imaging
  device}.  In the optical realm, the compressive ``single-pixel
camera'' has gained a lot of interest in research and practical
applications in the last decade
\cycite{baraniuk_compressive_2007}~\cycite{duarte_single-pixel_2008}.
Such an approach allows to image a scene using only one single
detector (instead of a 2D detector array as is usual in most optical
systems, for example in digital cameras).
In particular for imaging
applications outside the visible spectrum, there are many reasons why
choosing only a single detector might be advantageous: detectors might
be very expensive and building an array of them might not be
economically reasonable, or building a detector array might not be
technically feasible due to the degree of miniaturization that would
be needed for a practical application.
A concrete example here is terahertz imaging which is of special
interest in many different applications as terahertz radiation is
non-ionizing and at the same time able to penetrate many
non-conducting materials.

\begin{figure}
            \centering
            \includegraphics[width=1.0\linewidth]{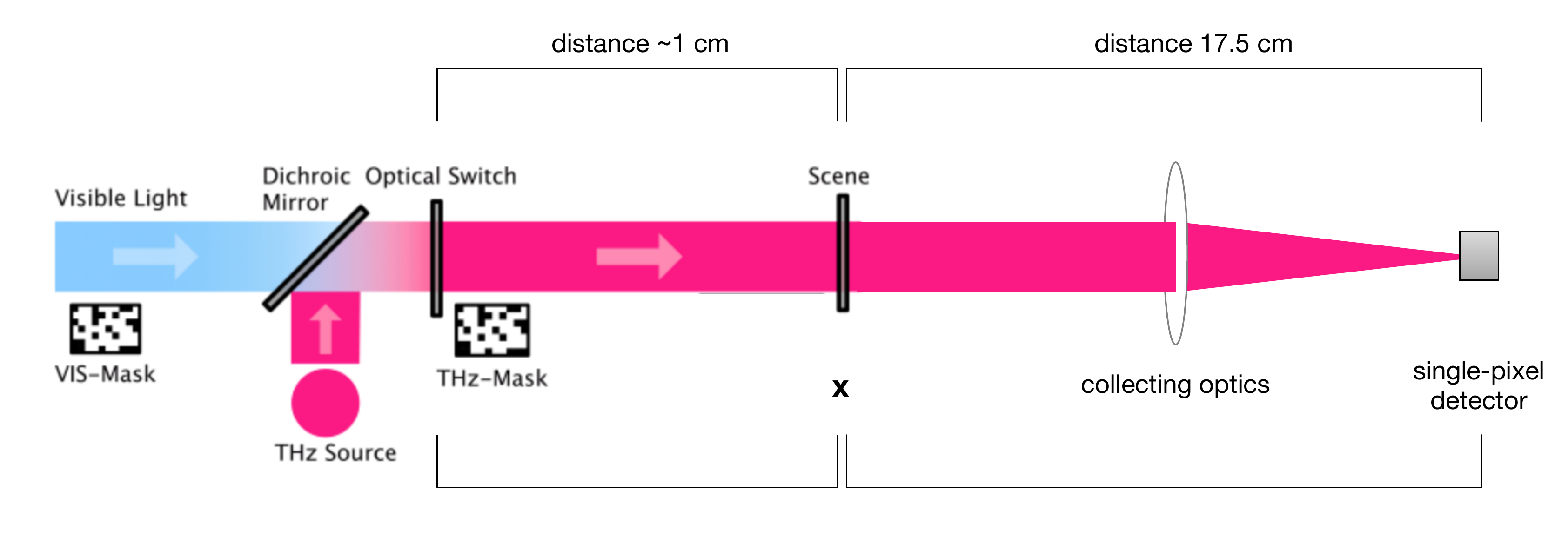}
            \caption{Schematic view of the experimental setup. The optical switch is controlled by visible light and allows the transmission of the terahertz radiation at the parts of the mask which are set to \texttt{1}, thereby imposing the pattern on the transmitted terahertz radiation. The terahertz pattern is propagated to the scene $\mathbf{x}$ and its transmission is collected using a collecting optics that focuses the radiation onto a singe detector cell. Figure adapted from work by Augustin et al. \cycite{augustin_optically_2017}.}
            \label{fig:thz:system_setup}
    \end{figure}

Our sketched application follows the system setup from \cycite{augustin_optically_2017}.  We are interested in
reconstructing the transmission of a scene illuminated with terahertz
radiation (at a wavelength of $0.856$nm, which equals to approximately
$0.35$ THz). However, since we restrict ourselves to only have a
single detector cell, we illuminate the scene with a \emph{random} but
\emph{known} radiation pattern and collect the transmission radiation
through a collecting optics (e.g. a lens) which focuses the
transmitted radiation into a single detector cell that is able to
measure the intensity (i.e. squared amplitude) of the incoming
radiation. This process is repeated with multiple different patterns
to obtain multiple measurements, which are then used to reconstruct
the original signal. Figure~\ref{fig:thz:system_setup} shows a schematic view
of the experimental setup.

The random radiation patterns are achieved by the usage of so-called
\emph{masks} applied by a spatial light modulator (also called optical
switch), a special device that modulates a radiation beam in a way
that it only allows radiation to pass through at certain selectable
areas (which are defined by the masks). The masks are 2-dimensional
patterns represented as binary vectors $\mathbf{a}_i \in\{0,1\}^{n}$. 

    Every electromagnetic wave is subject to diffraction effects when
    propagating through space after hitting an obstacle or propagating
    through an aperture in the size similar to its wavelength. In the
    optical regime these diffraction effects are negligible for many
    practical applications due to the extremely short wavelength of
    the visible light. However, in our setup, we will assume our target of
    interest to be of a size that is similar to the wavelength,
    especially will we assume that the pixel size of the image of the
    scene that we want to recover is approximately equal to the
    wavelength of the radiation.
    
    Technically, this means that we will have to model the diffraction
    effects taking place between the spatial light modulator and the
    scene and between the scene and the detector. Diffraction effects
    between the terahertz source and the spatial light modulator can
    be neglected as the radiation can be seen as coherent.  We use the
    ``Discrete Diffraction Transformation'', introduced in
    \cycite{katkovnik_backward_2009} and
    \cycite{katkovnik_discrete_2008}, to approximate the diffraction
    by complex matrices depending on the wavelength, the propagation
    distance and the pixel sizes at the planes before and after the
    propagation. For THz imaging via phase retrieval this has
    already been investigated in \cycite{burger_reconstruction_2019}.

    More precisely, we model diffraction effects between the spatial
    light \textbf{M}odulator and the \textbf{S}cene using a
    diffraction matrix
    $D_{\mathrm{M} \rightarrow \mathrm{S}} \in \mathbb{C}^{n \times
      n}$ generated according to the simplified construction method in
    \cycite{katkovnik_backward_2009}
    by assuming a propagation distance of 1cm (according to the system setup in Figure \ref{fig:thz:system_setup}), and $28 \times 28$ quadratic pixels of edge size 0.5mm (both before and after the propagation) at a wavelength of $0.856\cdot 10^{-3}$m. We define a second diffraction matrix, modeling the effects between the \textbf{S}cene and the \textbf{D}etector, $D_{\mathrm{S} \rightarrow \mathrm{D}} \in \mathbb{C}^{n \times n}$ and will generate it analogously assuming a propagation distance of $17.5$cm.

\subsection{Diffraction Imaging Measurement Model}
    
For our simulation, we will model the measurement at the detector as
follows: a uniform illumination $[1,\dots,1]^\top$ hits the spatial
light modulator which applies the mask
$\mathbf{a}_i$. After that, the wave propagates from
the spatial light modulator to the scene while being subject to
diffraction $D_{\mathrm{M} \rightarrow \mathrm{S}}$ before it hits the
scene $\mathbf{x}$ and propagates further from the
scene to the detector being subject to diffraction
$D_{\mathrm{S} \rightarrow \mathrm{D}}$. At the detector it is summed
up and its intensity is measured, leading to the following (noise-free) signal model:
\begin{equation}
  \begin{split}
    y_i & =  | \sum_{j=1}^n \left(  D_{\mathrm{S} \rightarrow \mathrm{D}} \mathrm{diag}(\mathbf{x}) D_{\mathrm{M} \rightarrow \mathrm{S}} \mathbf{a}_i\right)_j |^2 \\
    &=\left|\langle \overline{D_{\mathrm{M} \rightarrow \mathrm{S}}} \text{diag}(\mathbf{a}_i) D_{\mathrm{S}\rightarrow \mathrm{D}}^\mathrm{H}[1,\dots,1]^\top, \mathbf{x} \rangle\right|^2 \\
    &=:\left|\langle \tilde{\mathbf{a}}_i,\mathbf{x}\rangle\right|^2
  \end{split}
\end{equation}
for the phase retrieval optimization problem \eqref{eq:solving:drgd},
i.e., $\tilde{\mathbf{a}}_i$ is the $i$th row of the complex-valued measurement
matrix $\mathbf{A}$ in \eqref{eq:solving:drgd}.
    
\begin{figure}
            \centering
            \includegraphics[width=1.0\linewidth]{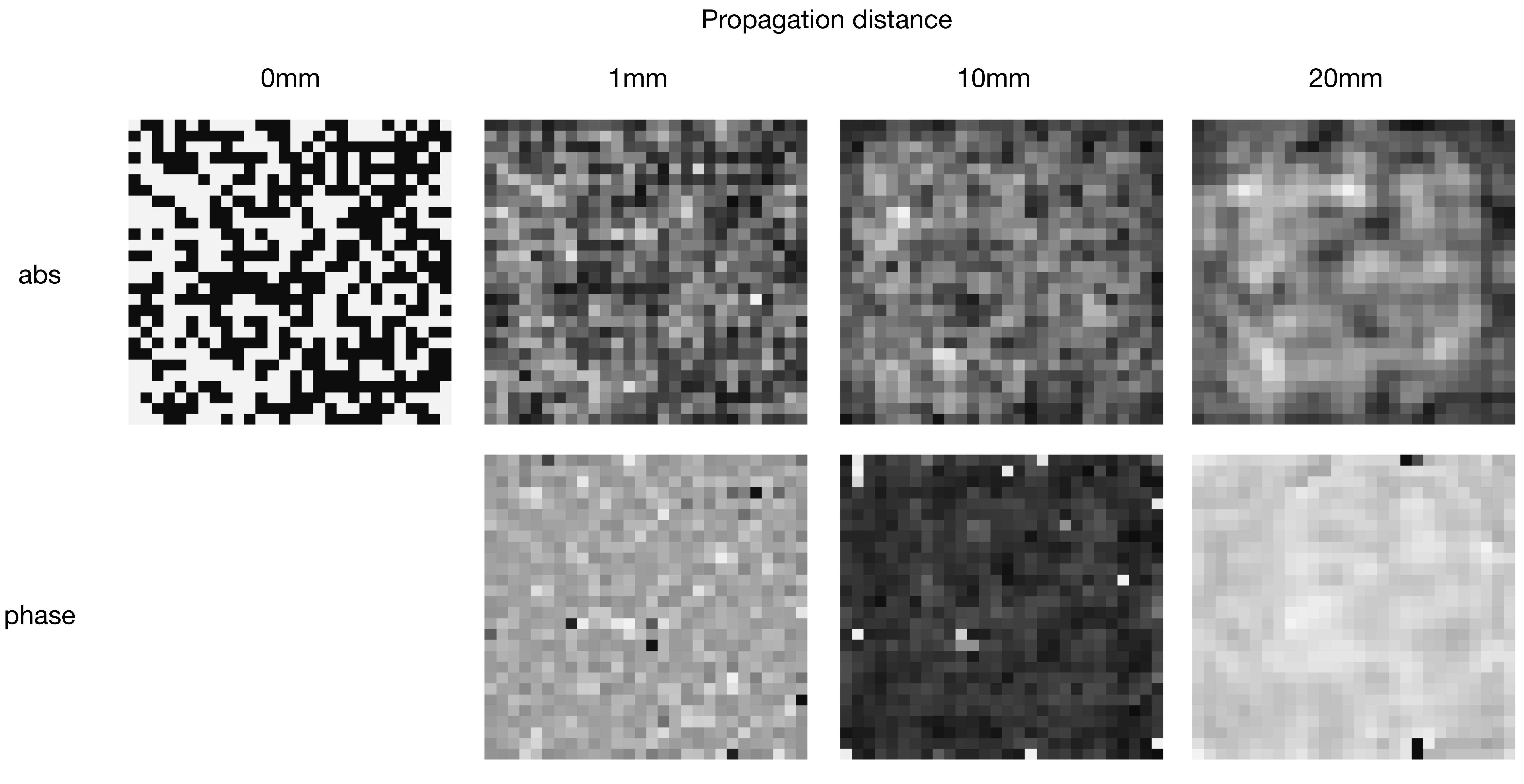}
            \caption{Commanded Bernoulli mask (0mm) and degraded masks after 1mm, 10mm and 20mm of simulated free-space propagation.}
            \label{fig:thz:mask_degradation}
    \end{figure}

Note that this imaging problem is especially sensitive to changes of
the distance between the spatial light modulator and the scene
$D_{\mathrm{M} \rightarrow \mathrm{S}}$ (also referred to as the
\emph{stand-off distance}), because the masks commanded at the spatial
light modulator drastically degrade while propagating to the scene
(see Figure~\ref{fig:thz:mask_degradation}). This is caused by the
diffraction matrix $D_{\mathrm{M} \rightarrow \mathrm{S}}$ losing rank
with increasing propagation distance (an effect that is covered in
more detail in \cycite{katkovnik_backward_2009}). This results in a
blurring effect, which, depending on the distance, can be up to a
degree that the original signal can no longer be recovered. We will
therefore investigate the reconstruction quality with respect to
sensitivity to changes in the distance between the spatial light
modulator and the scene for simulated data. A similar problem has been
investigated experimentally in \cycite{augustin_terahertz_2019}.
    
\subsection{Sensitivity Analysis}
\label{section:solving:sensitivity}
    
\begin{figure}
        \centering
        \includegraphics[width=0.95\linewidth]{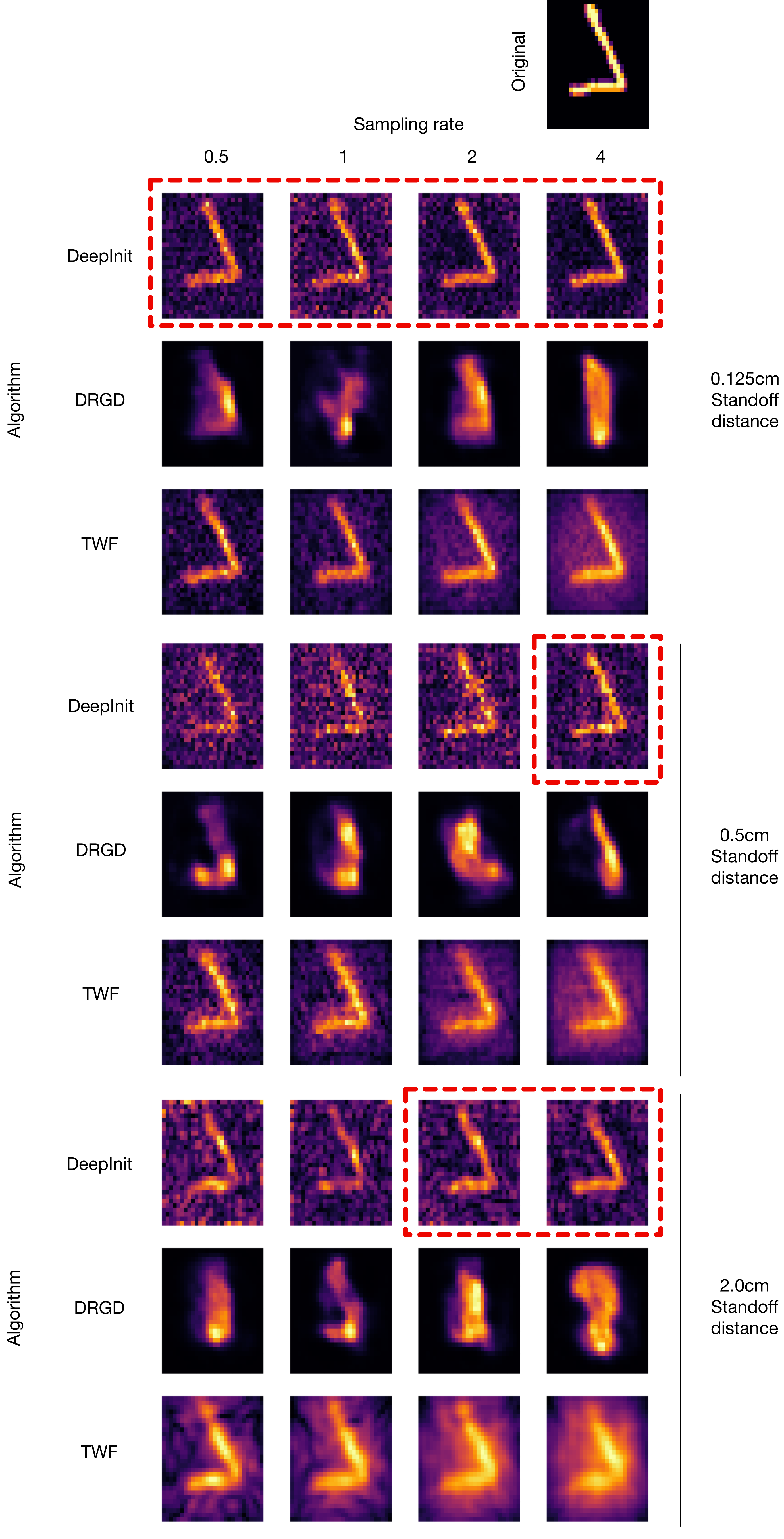}
        \caption{Results of the reconstruction process for a selected MNIST test image for selected sampling rates at 0.125cm, 0.5cm and 2cm stand-off distances. Important results are highlighted with a dashed red box.}
        \label{fig:results:practical:visual:mnist}
    \end{figure}

\begin{figure}
        \centering
        \includegraphics[width=0.935\linewidth]{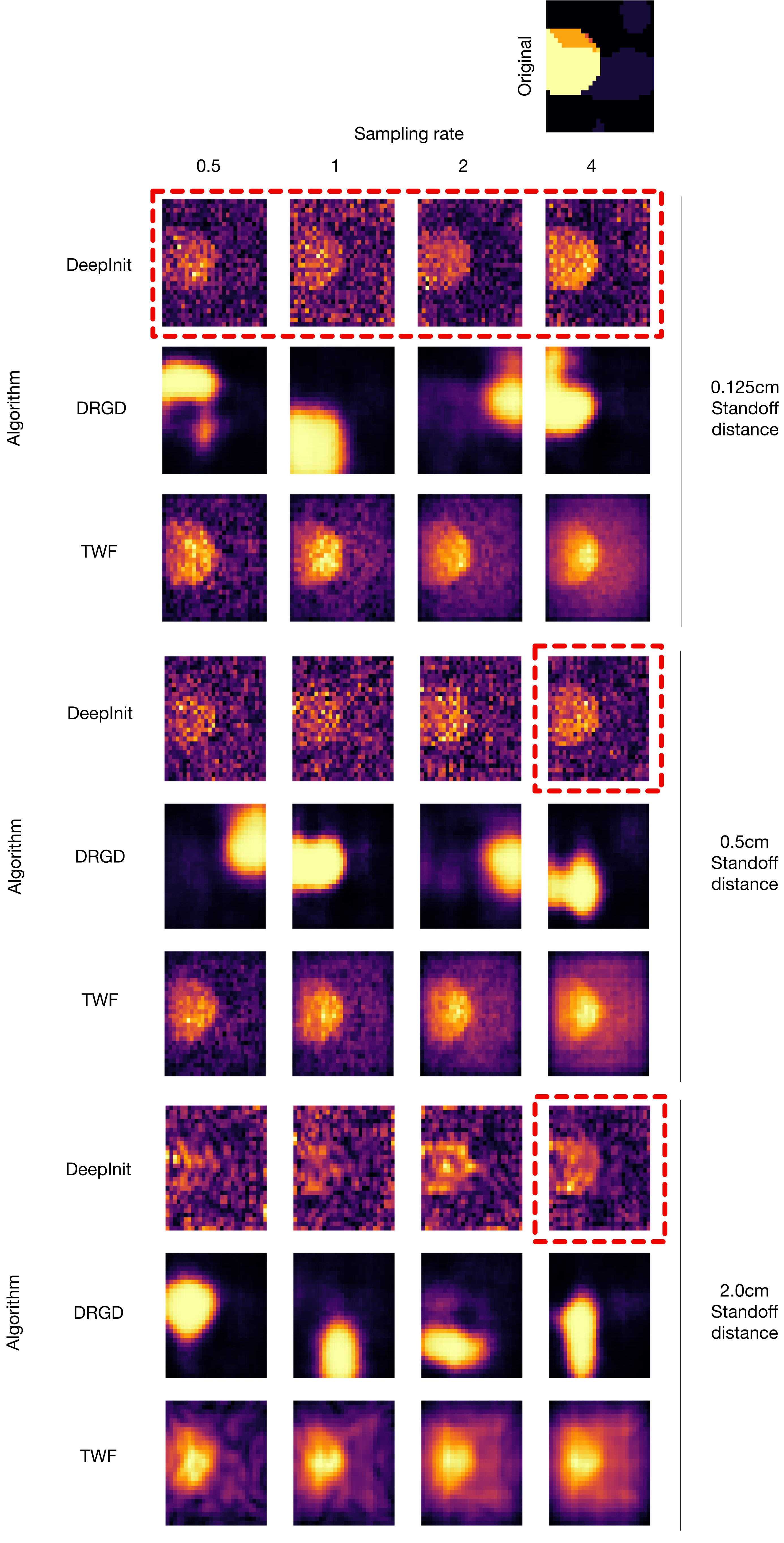}
        \caption{Results of the reconstruction process for a sample from the Shepp-Logan dataset for selected sampling rates at 0.125cm, 0.5cm and 2cm stand-off distances. Important results are highlighted with a dashed red box.}
        \label{fig:results:practical:visual:shepplogan}
    \end{figure}

We investigate the sensitivity of the reconstruction quality for
different stand-off distances and sampling rates $\frac{m}{n}$. Our
evaluation will be done for both traditional (Truncated Wirtinger
Flow) as well as deep generative prior-supported reconstruction
algorithms (Deep Regularized Gradient Descent and DeepInit Phase
Retrieval, both using the variational autoencoders defined in Sections
\ref{section:solving:mnist_dataset} and
\ref{section:solving:shepplogan_dataset} as their underlying
generative models) and will be executed on both the MNIST dataset as
well as on the synthetic Shepp-Logan dataset. To obtain reasonable
results for the Truncated Wirtinger Flow algorithm, we use parameters
($a_z^{\mathrm{lb}}=0.001$, $a_z^{\mathrm{ub}} = 500$) different to
the usual defaults in \cycite{chen_solving_2015}.
Figures~\ref{fig:results:practical:visual:mnist} and
\ref{fig:results:practical:visual:shepplogan} visually show the
results of the reconstruction process for selected MNIST and
Shepp-Logan samples for stand-off distances between $0.00125$m and
$0.08$m.

 approaches and can prove valuable
in real-world physical image reconstruction problems where diffraction
effects occur.


\section{Conclusion}
    
This work explored how deep generative models can support solving phase retrieval problems.    
Recent works approach solutions in the range of a trained generator
network, but suffer from poor reconstruction quality when the
generator is not properly able to model the signal domain.
\emph{DeepInit Phase Retrieval} also incorporates signal domain
information using deep generative priors but does not suffer from
reconstruction quality degradation caused by generator model
error. This is because the data prior is used only during the
initialization while the actual reconstruction is performed using
classical algorithms like Randomized Kaczmarz iterations. Our work
empirically shows that DeepInit Phase Retrieval achieves better
reconstruction quality than provided by traditional methods at
sampling rates below $4$ and even comes with improved runtime compared
to other gradient descent methods.
For the practically motivated application in terahertz single-pixel
imaging, we experimentally showed that DeepInit Phase Retrieval
achieves reconstruction quality that is superior to Truncated
Wirtinger Flow, indicating that it is well suited for real-world scenarios
where diffraction plays an important role.
\subsubsection*{Acknowledgements}
We thank Sven Augustin and Lukas Nickel. PJ has been supported by DFG grant JU 2795/3.

\bibliography{arxiv}

\end{document}